\documentclass[twocolumn,english,showpacs,floatfix]{revtex4}

\usepackage{amsmath}
\usepackage{bbold}
\usepackage{graphicx}
\usepackage{amsfonts}
\usepackage{amssymb}
\usepackage[latin9]{inputenc}
\usepackage{babel}
\usepackage{latexsym}
\usepackage{dcolumn}

\newcommand{\be}{\begin{equation}}
\newcommand{\ee}{\end{equation}}

\begin{document}

\title{The signature of dark energy perturbations 
in galaxy cluster surveys} 

\author{L. Raul Abramo}
\email{abramo@fma.if.usp.br}
\author{Ronaldo C. Batista}
\affiliation{Instituto de F\'{\i}sica, Universidade de S\~ao Paulo, 
CP 66318, 05315-970, S\~ao Paulo, Brazil} 

\author{Rog\'erio Rosenfeld}
\affiliation{Instituto de F\'{\i}sica Te\'orica, Universidade Estadual Paulista,
R. Pamplona 145, 01405-900, S\~ao Paulo, Brazil}

\begin{abstract}
All models of dynamical dark energy possess fluctuations, which
affect the number of galaxy clusters in the Universe.
We have studied the impact of dark energy clustering on
the number of clusters using a generalization of the spherical 
collapse model and the Press-Schechter formalism. 
Our statistical analysis
is performed in a 7-parameter space using the Fisher matrix method,
for several hypothetical Sunyaev-Zel'dovich and weak lensing 
(shear maps) surveys.
In some scenarios, the impact of these fluctuations is large
enough that their effect could already be detected by existing 
instruments such as the South Pole Telescope, when its data
is combined with WMAP and SDSS.
Future observations could go much further and 
probe the nature of dark energy by distinguishing between 
different models on the basis of their perturbations, not only their
expansion histories.
\end{abstract}

\pacs{98.80.C9; 98.80.-k; 98.65.Cw}

\maketitle


\noindent{\it Introduction} -- 
Over the past few years evidence for the
accelerated expansion of the Universe has become 
overwhelming: besides supernovas
\cite{Perlmutter:1998np,Riess:1998cb,Riess:2004nr,Wood-Vasey:2007jb},
every well-measured cosmological observable seems to support 
the conclusion that we no longer live in a decelerated, matter-dominated
Einstein-de Sitter Universe 
\cite{Astier:2005qq,BAO,Spergel:2006hy,Wang:2006ts,Wright:2007vr}. 
The search for the cause of this acceleration, 
whether it is yet another dark component (dark energy) 
or if it is due to some modification of gravity which makes it repulsive
at large scales
\cite{Dvali:2000hr,Carroll:2003wy,Carroll:2004de,Jain:2007yk}, has become 
one of the most pressing  questions of our times.

The properties of dark energy, and its effect on cosmological observables,
are commonly parametrized in terms of an equation of state 
$w_{de}=p_{de}/\rho_{de}$ with a simple, dependence on redshift,
$w_{de}(z)=w_0 + w_a z/(1+z)$ \cite{Chevallier:2000qy,Albrecht:2006um}.
Substantial observational efforts are underway \cite{Ruhl:2004kv,Abbott:2005bi,Ivezic:2008fe}, 
and many more are being planned
\cite{Aldering:2004ak,Refregier:2006vt}, in order
to determine such parameters.

However, getting the equation of state even with excellent precision, 
in whatever parametrization, would still tell us close to nothing about the nature of dark energy \cite{Uzan:2006mf}, 
or if the acceleration is in fact due to modified gravity. 
One way of answering those questions is to look for 
evidence of dark energy perturbations in cosmological 
observables -- evidence which must exist if dark energy is 
anything but the Cosmological Constant \cite{Coble:1996te}.
Notice that the signal of dark energy perturbations should not be confused
with the influence of smooth dark energy on the growth function
of dark matter inhomogeneities, which is another important,
but distinct effect of dark energy upon structure
formation \cite{Linder:2003dr,Linder:2005in,Ballesteros:2008qk}.

Unfortunately, searching for imprints of dark energy fluctuations 
can sometimes be underwhelming: since dark energy dominates 
only at late times, its impact on the cosmic 
microwave background (CMB) arises only through the integrated 
Sachs-Wolfe effect, which in turn makes the signal 
both small, and limited to the largest
scales where cosmic variance defeats the statistical purpose of the data 
\cite{Erickson:2001bq,DeDeo:2003te,Abramo:2004ji,Hu:2004yd}.

Dark energy fluctuations also affect the power spectrum
$P(k)$ \cite{Caldwell:1997ii,Chiba:1998de,Erickson:2001bq,
DeDeo:2003te,Uzan:2006mf}. 
In particular, Hu \cite{Hu:2001fb} used an effective 
description for the dark energy pressure perturbations in order to estimate 
their impact on some cosmological observables. 
In that approximation, the pressure perturbation $\delta p_{de}
= s_{\rm eff} \, \delta \rho_{de}$, where $s_{\rm eff}$ is an {\it effective}
sound speed squared and $\delta \rho_{de}$ is the energy 
density perturbation of dark energy
 \footnote{This is {\em not} an approximation only if one works in the comoving
gauge \cite{Hu:2001fb}, in which case
obviously the Einstein equations inherit
the gauge terms.}. 
This framework is not ideal, but it 
is extremely useful if our goal is to enlarge the class of models
we want to constrain, just like any given parametrization of the 
equation of state is not meant to represent
actual models, but to simplify the task of assessing the power
of observations to constrain those models.
Along these lines, e.g. Takada \cite{Takada:2006xs} regarded 
$s_{\rm eff}$ as a 
free parameter and concluded that measurements of the CMB and 
power spectrum could detect the effects of dark energy 
inhomogeneities, but only if the effective sound speed was very 
small, which would make it behave almost like dust on 
the scales probed
by the CMB and $P(k)$.

That leaves the nonlinear regime of structure formation (collapsed 
structures, or halos)
as one of the few venues left where we could expect the 
influence of dark energy
to be measurable. As the initially overdense regions attract the
surrounding matter and experience gravitational collapse, the
steep gravitational potentials at the center of the dark matter halos 
deform the distribution of dark energy.
And since pressure is a key ingredient in halo formation, dark energy 
could have an enhanced effect on the masses, times of formation
and relative abundances of halos 
\cite{Mota:2004pa,Nunes:2004wn,Manera:2005ct,Nunes:2005fn,Maor:2005hq,Liberato:2006un,Abramo:2007iu,Abramo:2007mv,Abramo:2008ip}
as well as voids \cite{Dutta:2006pn,Mota:2007zn}. Notice that both 
$P(k)$ and galaxy cluster counts are potentially sensitive tests of
dark energy perturbations. However, while the former tests
the long-range, infra-red properties of the underlying theory of 
dark energy in the linear regime, the latter is a test of the 
small-scale, ultraviolet, nonlinear regime of the same unknown 
theory. They are, therefore, complementary rather than competing 
tests of the nature of dark energy -- see also
\cite{Schmidt:2008tn} on using cluster counts to constrain 
models of modified gravity.


\vskip 0.3cm

\noindent{\it Methodology} --
Counts of galaxy clusters have long been regarded as a crucial cosmological observable
\cite{Viana:1995yv,Bahcall:1997ia}, in particular for
purposes of constraining dark energy models
\cite{Wang:1998gt,Haiman:2000bw,Weller:2001gk,Battye:2003bm,Basilakos:2003bi,Lokas:2003cj,Wang:2004pk}.
In the present work we study 
the sensitivity of cluster counts to dark
energy perturbations.
We employ a set of seven cosmological parameters that we
allow to vary: $h$ (the value of the Hubble expansion rate, where 
$H_0 = 100$ $h$ km s$^{-1}$ Mpc$^{-1}$), 
$\Omega_m$ (the amount of matter relative to the critical density), 
$\Omega_b$ (density ratio of baryons), $\sigma_8$ (rms matter fluctuations
on scales of 8 h$^{-1}$ Mpc), $w_0$, $w_a$ and $s_{\rm eff}$ (we
assume flatness throughout.)
The abundance of halos is then estimated with
the Press-Schechter formalism \cite{Press:1973iz},
using the top-hat spherical collapse
model  \cite{Gunn:1972sv} to compute the critical density at the time of
collapse -- see also \cite{Viana:1995yv,Fosalba:1997tn,Percival:2001nv,Mota:2004pa}, and, in particular, \cite{Abramo:2007iu,Abramo:2008ip}.

Our statistical analysis relies on the Fisher
matrix, which was computed in this 7-dimensional parameter space
for several hypothetical Sunyaev-Zel'dovich (SZ) and weak lensing 
(WL) surveys of galaxy clusters,
and a suitable fiducial dynamical dark energy (DDE) model
with $w_0 = -1.1$ and $w_a=0.5$. We have checked
that the impact of dark energy clustering scales
like $| 1+w|$, which is natural since in the $\Lambda$CDM limit
there are no dark energy perturbations.
The other fiducial values we used are $h=0.72$, 
$\Omega_b=0.05$, $\sigma_8=0.76$ and
$\Omega_m = 0.25$.
This model lies near the best-fit 
region in the parameter space of several recent joint analyses of 
cosmological observations \cite{Wang:2008zh,Vikhlinin:2008ym}. 
Notice that, even though our DDE model happens to experience
``phantom crossing'' ($w=-1$) at a redshift $z=0.25$,
there is no associated instability in the spherical collapse model,
hence the sensitivity to dark energy is not artificially enhanced in
this scenario compared to similar scenarios without phantom crossing
-- see, for instance, \cite{Hu:2004kh}.
Finally, for completeness we studied three scenarios 
of DDE, with fiducial values 
$s_{\rm eff}=0$,
$s_{\rm eff}=0.5$ and $s_{\rm eff}=-0.75$.

The main limitation to the sensitivity of number counts is
the halo mass below which clusters cannot be detected
with the given instrument and strategy. For the $i$-th bin
the number of observed clusters is given in terms of the mass
function as:
$$
N_i = \Delta \Omega 
\int_{z_i}^{z_{i+1}} dz
\int_{M_i}^{M_{i+1}} dM 
\, \theta[M-M_{min}(z)] \frac{dn}{dM dz} \; ,
$$
where $\Delta \Omega$ is the solid angle subtended by the survey.
The limiting mass $M_{min}$
is approximately constant in redshift for surveys which employ the SZ
effect, but in WL cluster surveys the limiting mass
grows with redshift due to the declining number of 
background galaxies \cite{Majumdar:2003mw,Marian:2006zp}.

We consider three hypothetical scenarios for these two types of surveys:
near-future (nf) ones, such as the SPT/DES \cite{Ruhl:2004kv}, 
which are supposed to cover about 4000 deg$^2$; 
future (f) ones, such as the LSST \cite{Ivezic:2008fe}, 
which will cover about half the sky (18.000 deg$^2$); 
and far-future (ff) ones, such as EUCLID/JDEM 
\cite{Refregier:2006vt,Albrecht:2009ct},
which will cover up to 75\% of the sky.

\begin{figure}
\vspace{0.5cm}
\includegraphics[width=4.2cm]{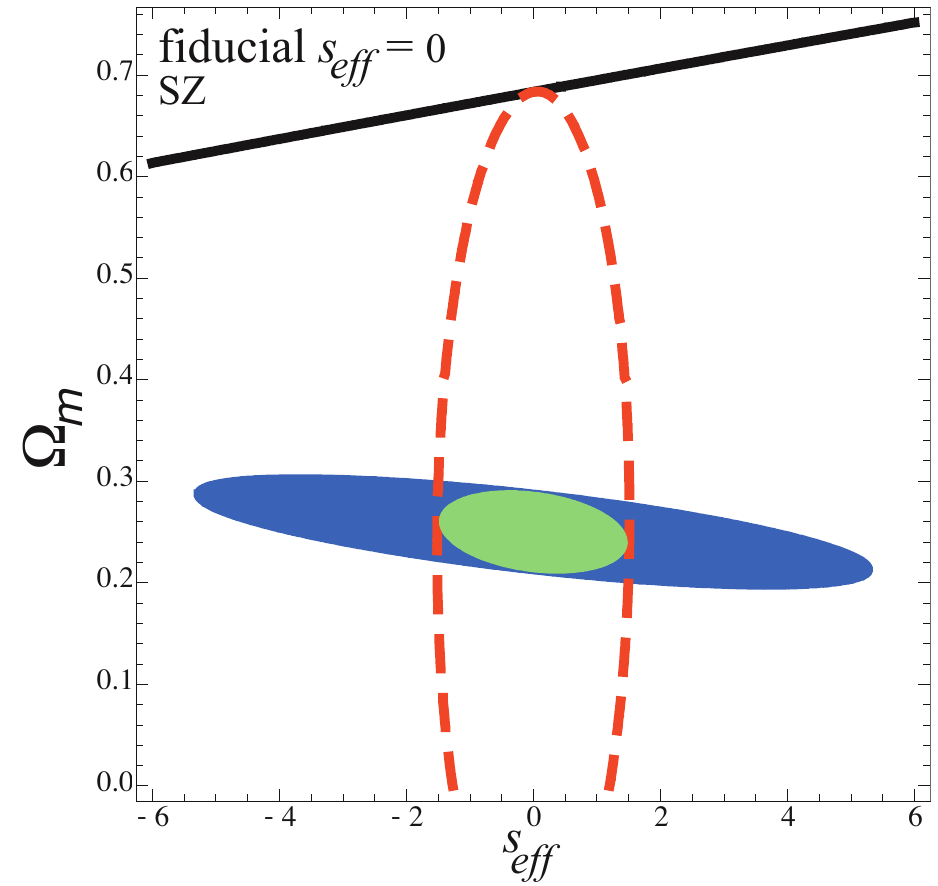}
\includegraphics[width=4.2cm]{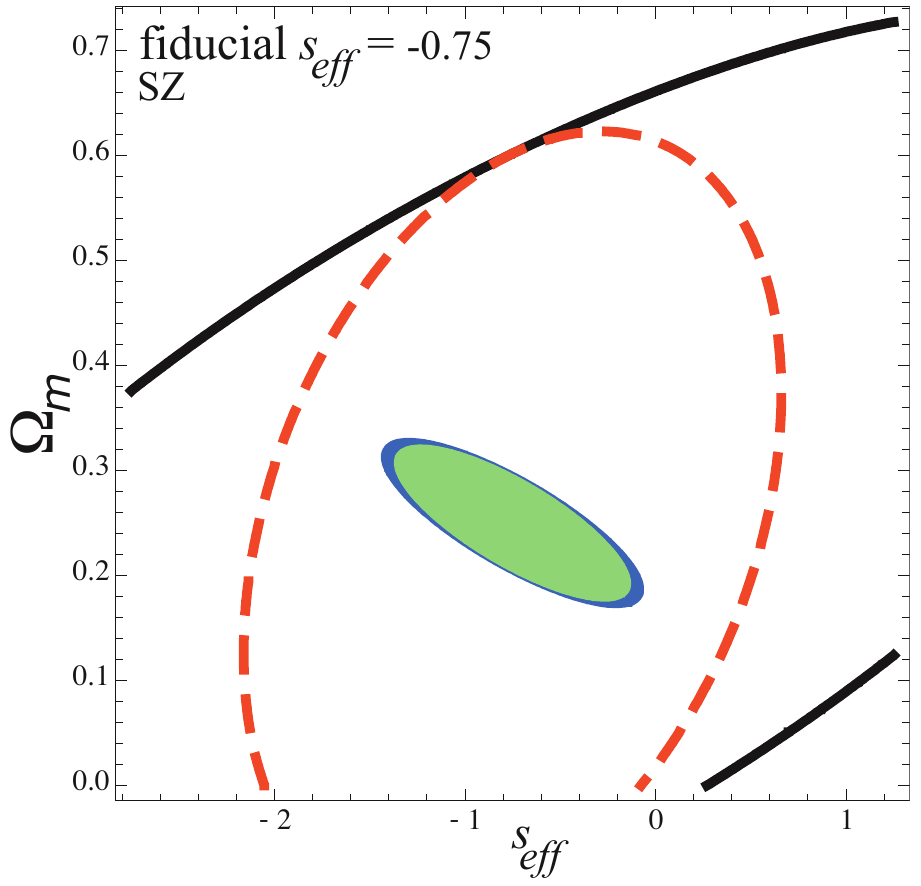}
\includegraphics[width=4.2cm]{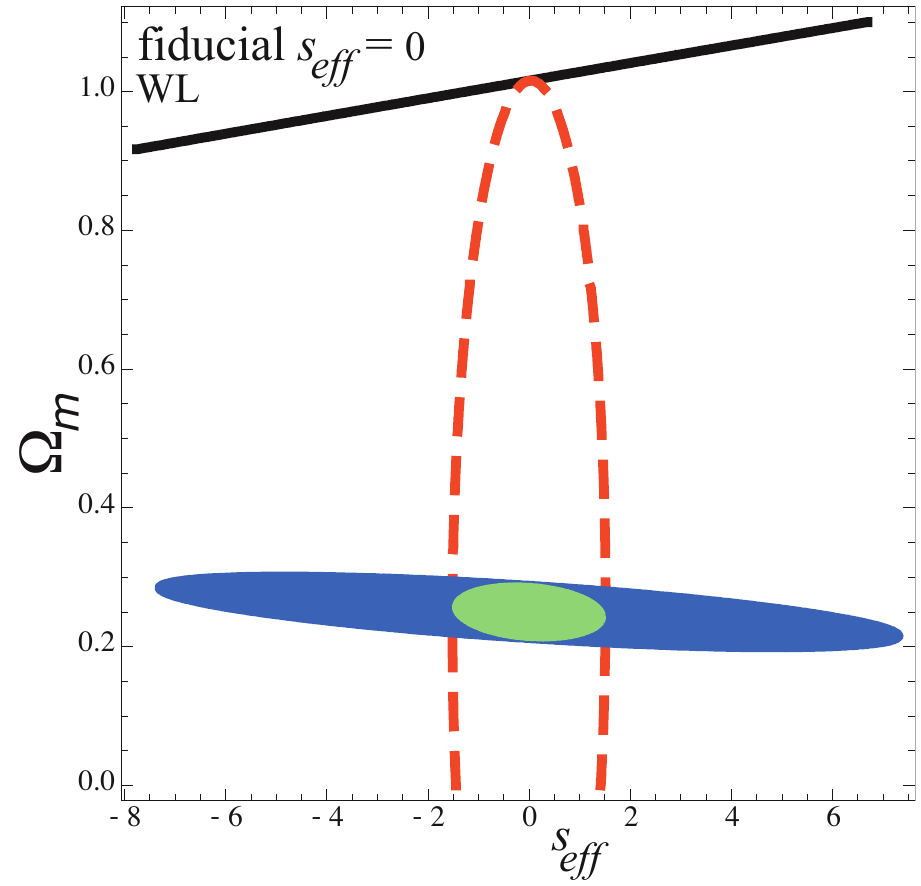}
\includegraphics[width=4.2cm]{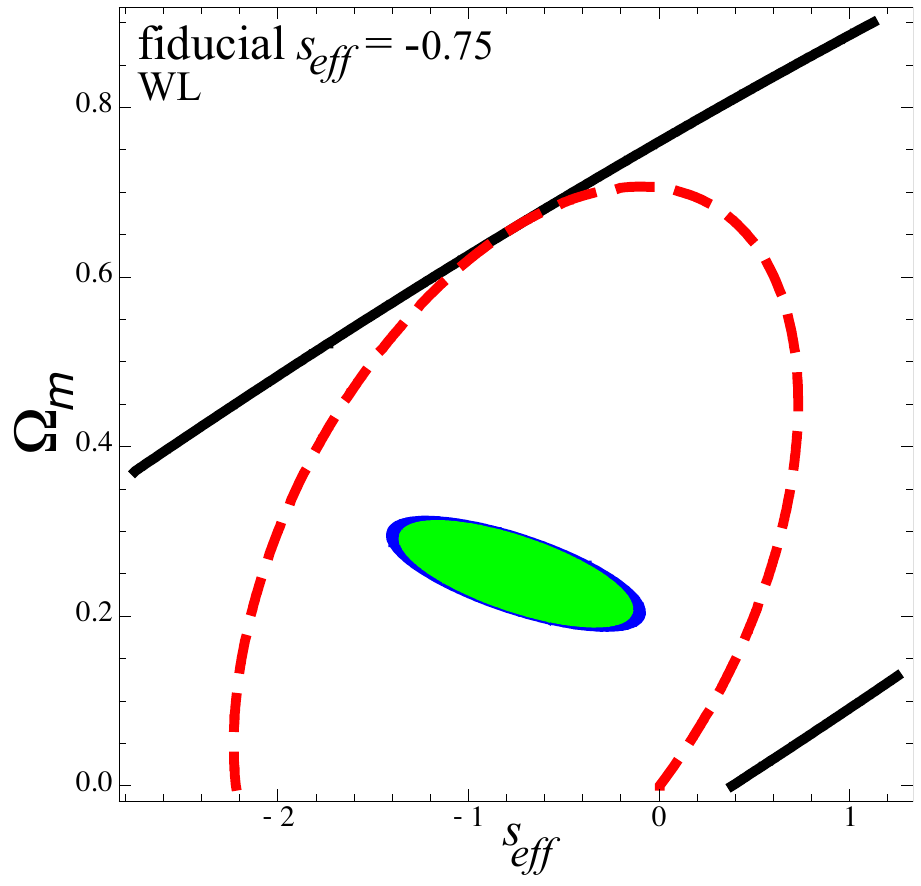}
\caption{Joint Constraints on $s_{eff}$ and $\Omega_m$ from
near-future (nf) surveys. The top panels show marginalized 
68\% C.L. limits from SZ-type surveys, 
and the bottom panels show the limits from WL-type surveys,
for the fiducial models with $s_{eff} = 0$ (left panels) and 
$s_{eff} = -0.75$ (right panels.) The thick solid (black in color version) 
lines correspond to cluster counts limits (68\% C.L., marginalized); 
the thick dashed (red) lines assume an additional weak prior on $s_{eff}$,
$\sigma_{s_{eff}}^2 = 1$; the 
dark (blue) ellipses correspond to adding the ``COSMO'' set of priors 
(see text) to the cluster Fisher matrix; 
and the smaller, light (green) ellipses correspond to adding 
both the weak prior on $s_{eff}$ and the COSMO priors.
}
\label{DDE_SZWL_ce2_Om$}
\end{figure}

\begin{figure}
\vspace{0.5cm}
\includegraphics[width=8cm]{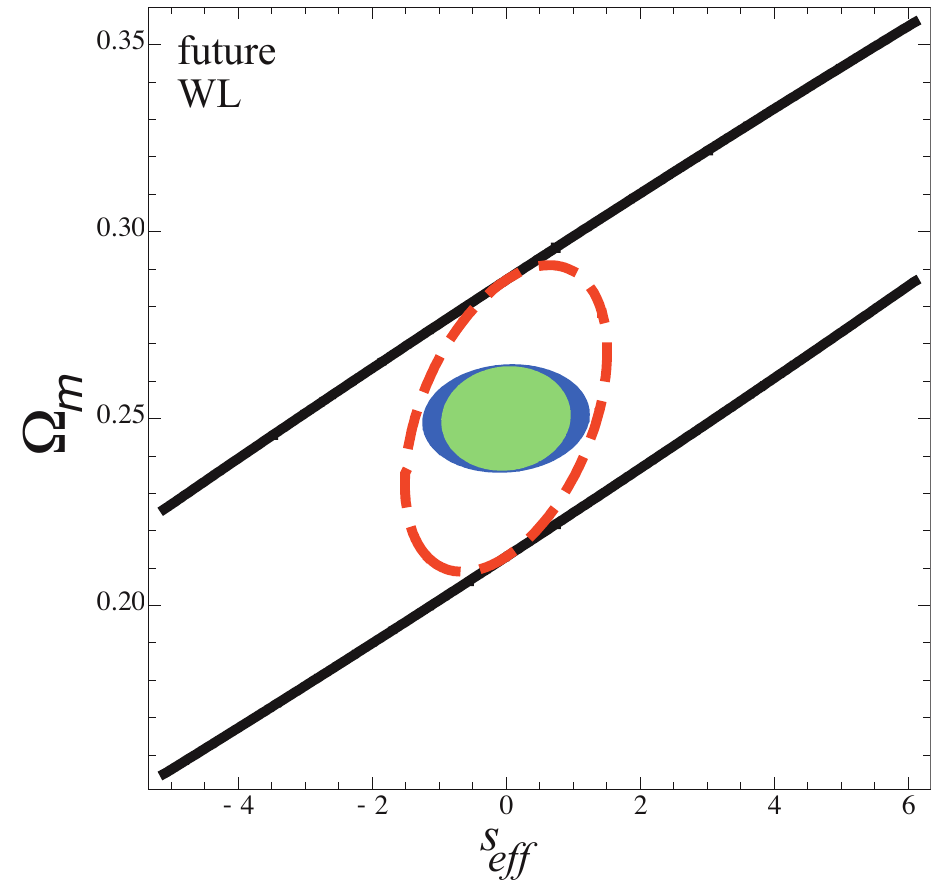}
\includegraphics[width=8cm]{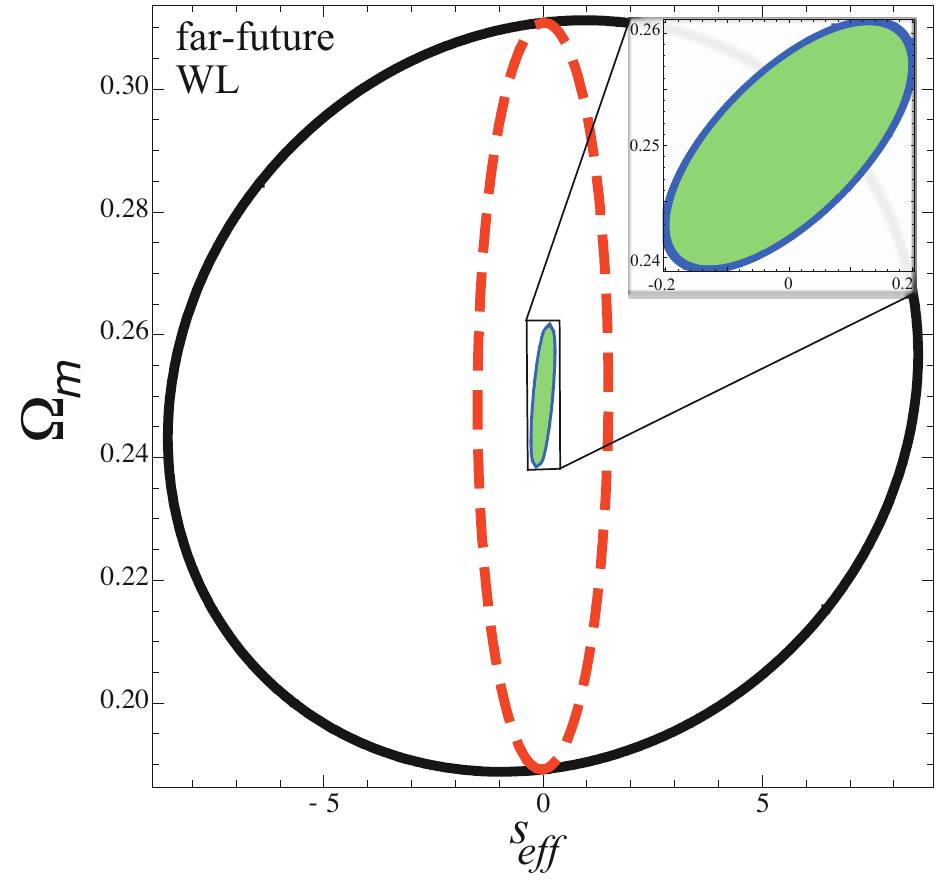}
\caption{Joint Constraints on $s_{eff}$ and $\Omega_m$ from
WL surveys in the fiducial model with $s_{eff} = 0$,
for future surveys (upper panel) and far-future surveys (lower.)
The colors are the same as in Fig. 1.
}
\label{DDE_SZ_ce2_Om_ce75$}
\end{figure}

We have defined limiting masses by a simple 
two-parameter function: $\log_{10} M_{\rm min}/(h^{-1} M_\odot)  
= \lambda + \beta z$. We have set $\lambda_{nf}^{SZ}=14.2$,
 $\lambda_f^{SZ}=14.1$ and $\lambda_{ff}^{SZ}=13.7$, 
 and $\beta^{SZ}=0$ in all cases -- which means we assume exactly flat
 selection functions for the SZ-like surveys, clearly a very 
 rough approximation.
For the WL surveys we have set
 $\lambda_{\rm nf}^{WL}=14.0$,
 $\lambda_{\rm f}^{WL}=13.5$ and $\lambda_{\rm ff}^{WL}=13.2$, and
$\beta^{WL}=0.6$ in all cases. 
With these limiting masses, we
obtained (for $\Lambda$CDM) 
total numbers of clusters of: 7000 and 4600 respectively
for the near-future
SZ and WL surveys; $\sim$ 280.000 for future surveys (both SZ and WL);
and $\sim 1.5 \times 10^6$ for far-future surveys (both SZ and WL).

We have used different binnings in mass and in redshift in each case: 
for near-future SZ and WL surveys we used
3 mass bins, and 10 redshift bins up to $z=1.5$; 
for future SZ and WL
surveys we used 5 mass bins and 15 redshift bins up to $z=1.5$;
and for the far-future SZ and WL
surveys we used 8 mass bins and 25 redshift bins up to $z=2.0$.
Our results do not change significantly if fewer mass bins
are used, and we
chose fat redshift bins in order to minimize 
sample (``2-halo'') variance 
\cite{Hu:2002we,Marian:2006zp,Takada:2007fq}.
In fact, we only consider Poisson (shot) noise in
the covariance of the number of detected halos:
$\langle (N_i - \bar{N}_i) (N_j - \bar{N_j}) \rangle = \delta_{ij} \bar{N}_i$,
where $N_i$ ($\bar{N}_i$) is the actual (expected) number of halos 
in the $i$-th bin.

The Fisher matrix is then given by: 
$$
F_{ab}= \sum_{i}
 \frac{1}{N_i} \, \frac{\partial N_i}{\partial \theta^a} \, \frac{\partial N_i}{\partial \theta^b} \; , 
 $$
 where $\theta^a$ are any one of the 7 cosmological parameters.
 If all other parameters are fixed, the 1$\sigma$ (68\% C.L.)
 limit on the parameter
 $\theta^a$ is given by $\sqrt{1/F_{aa}}$. If the other parameters are
 marginalized over (integrated out of the total PDF), 
 then the 1$\sigma$ limit on $\theta^a$ is given
 by $\sqrt{(F^{-1})_{aa}} \; .$
 
 We have also considered the effect of adding an external (``COSMO'')
 set of cosmological priors: WMAP 5y (through the shift parameter $R$) 
 \cite{Dunkley:2008ie}, BAO \cite{BAO}, HST \cite{Freedman:2000cf}
 and a baryon fraction $\Omega_b h^2 = 0.02273 \pm 0.00062$,
 also from WMAP 5y data \cite{Dunkley:2008ie}.


\vskip 0.3cm

\noindent{\it Results and Discussion} --
If the fiducial model is $\Lambda$CDM,
then naturally the sensitivity to perturbations (and to the sound speed) of dark energy vanishes. For the dynamical dark energy 
fiducial model we find that the impact of the dark energy perturbations
is dramatically different depending on the value of the dark energy
pressure perturbation. If the pressure is positive, zero or negative, 
then the impact on the mass function (and on the number counts) 
is respectively very small, small or very large (see Figs. 1-3.) This means
that the likelihood function is more peaked for smaller values of
the sound speed.

For positive dark energy pressure, our results are similar to what would
be expected from ``quintessence'' (canonical scalar field) models of 
dark energy, since in that case the Jeans length is
approximately equal to the Compton
wavelength of the dark energy field, $\lambda_J \simeq c_s/m_{de}$
($c_s=\sqrt{s_{eff}}$ is the sound speed.) In that case, dark energy 
clusters only on horizon scales, and its effect on cluster counts is minute.

For near-vanishing dark energy pressure (sound speed very small) our
results are similar to power-spectrum constraints like those by 
obtained by, e.g., Takada \cite{Takada:2006xs}. However, because the
typical Jeans lengths are still typically much larger than cluster scales, 
the limits from number counts are weaker than the limits from $P(k)$.

For negative pressures ($s_{eff}<0$) the sensitivity of galaxy cluster
counts to dark energy perturbations increases dramatically.
In fact, this would be true for the sensitivity of 
any large-scale structure observable (e.g., CMB and power spectrum),
if it were not for the fact that it is probably nonsensical to
consider negative sound speeds squared at the linear level. 
However, in collapsed regions and small scales there is nothing 
that forbids negative pressures \cite{Mota:2004pa,Abramo:2008ip} 
-- in fact, the bag model of hadronic physics is
one microphysical example that supports 
the use and interpretation of negative pressure in cosmology.

\begin{figure}
\vspace{0.5cm}
\includegraphics[width=4.2cm]{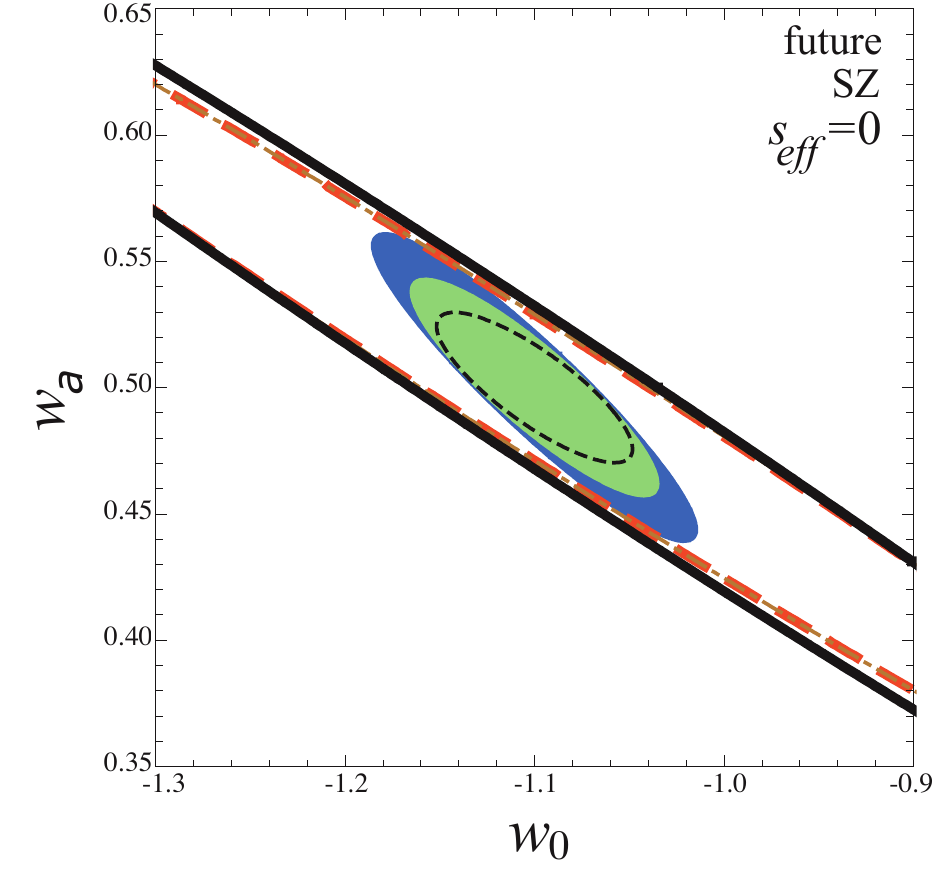}
\includegraphics[width=4.2cm]{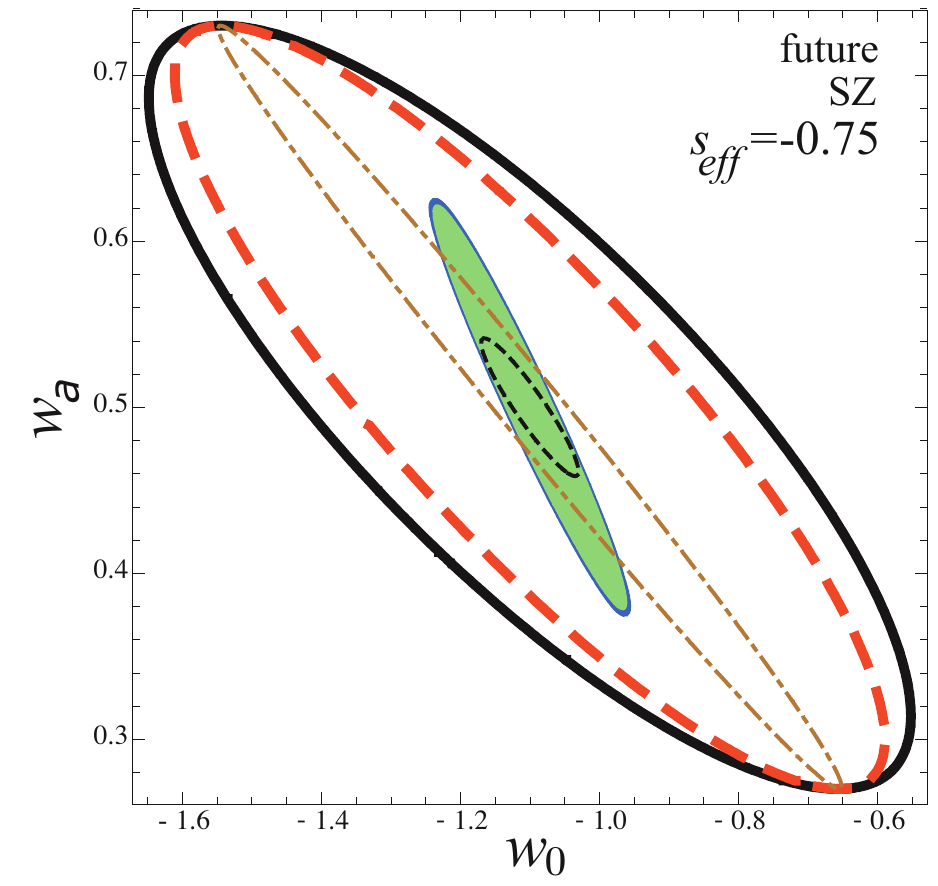}
\caption{Joint Constraints on $w_{0}$ and $w_a$ from
near-future SZ surveys, for the fiducial models with 
$s_{eff} = 0$ (left) and $s_{eff} = -0.75$ (right.)
Legends for the thick solid and dashed lines, as well as for
the ellipses, are identical to the previous figures.
In addition, computing the 68\% C.L. limits {\em without} the dark energy
clustering parameter $s_{eff}$ leads to either 
the thin dashed (black) or dot-dashed (brown) lines, whether or
not one includes the COSMO set of priors.
}
\label{DDE_SZ_w0wa$}
\end{figure}

\begin{figure}
\vspace{0.5cm}
\includegraphics[width=4cm]{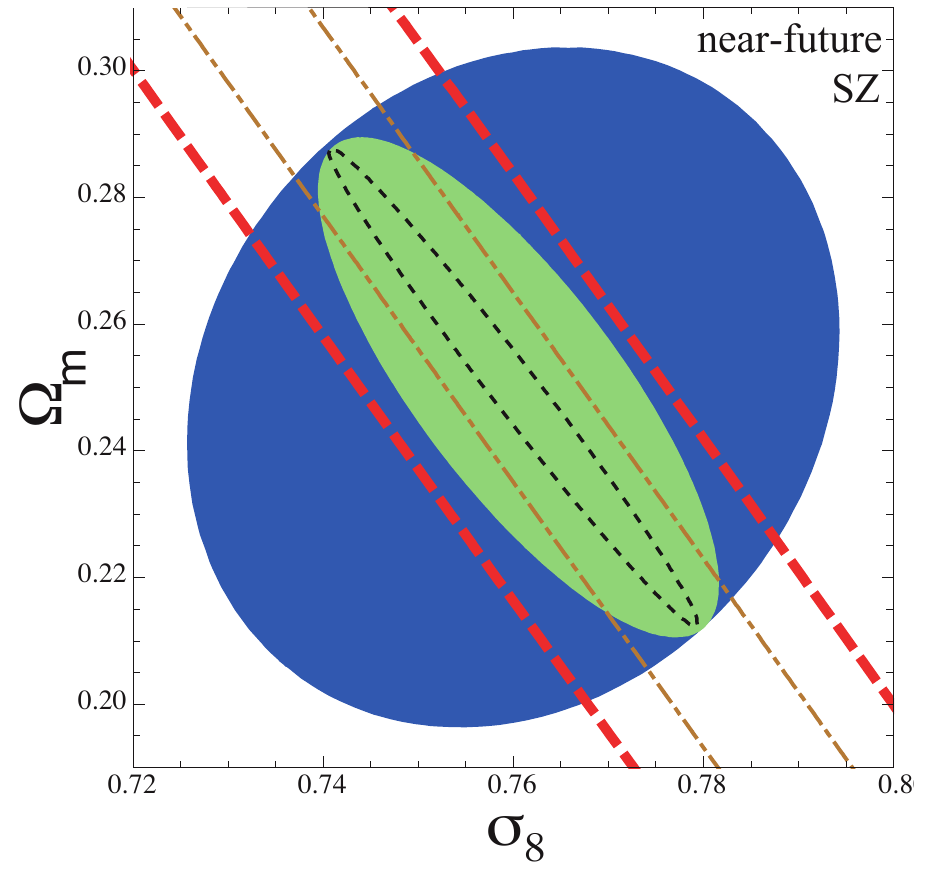}
\includegraphics[width=4cm]{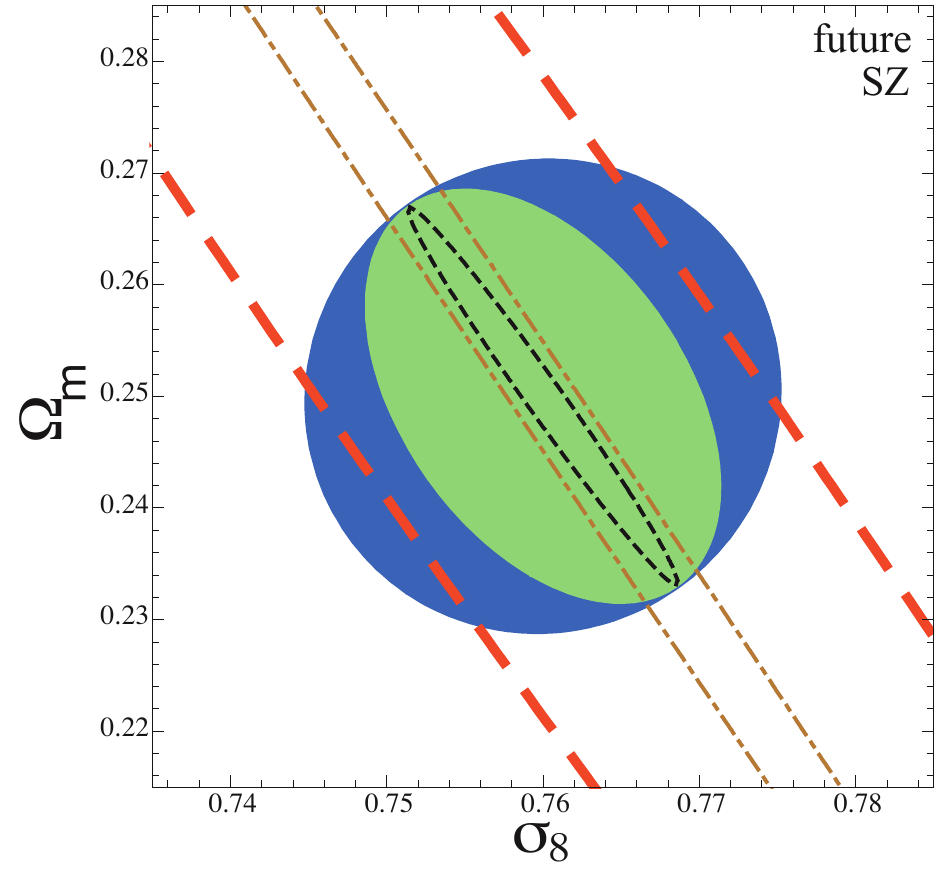}
\includegraphics[width=4cm]{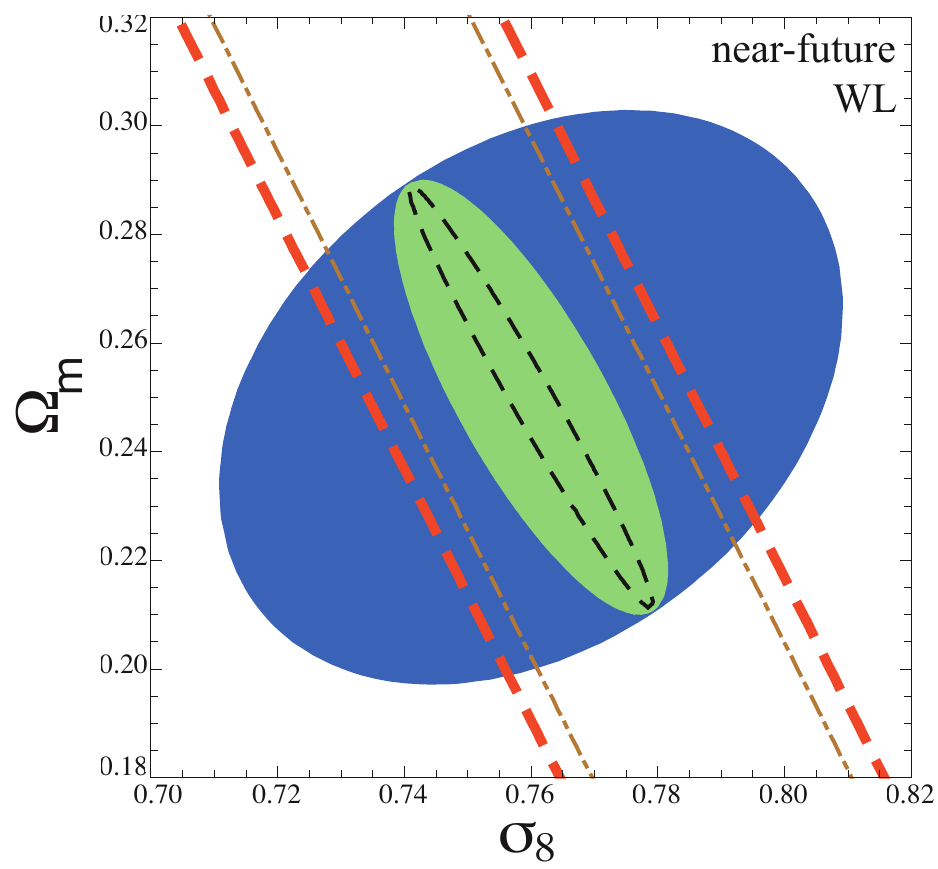}
\includegraphics[width=4cm]{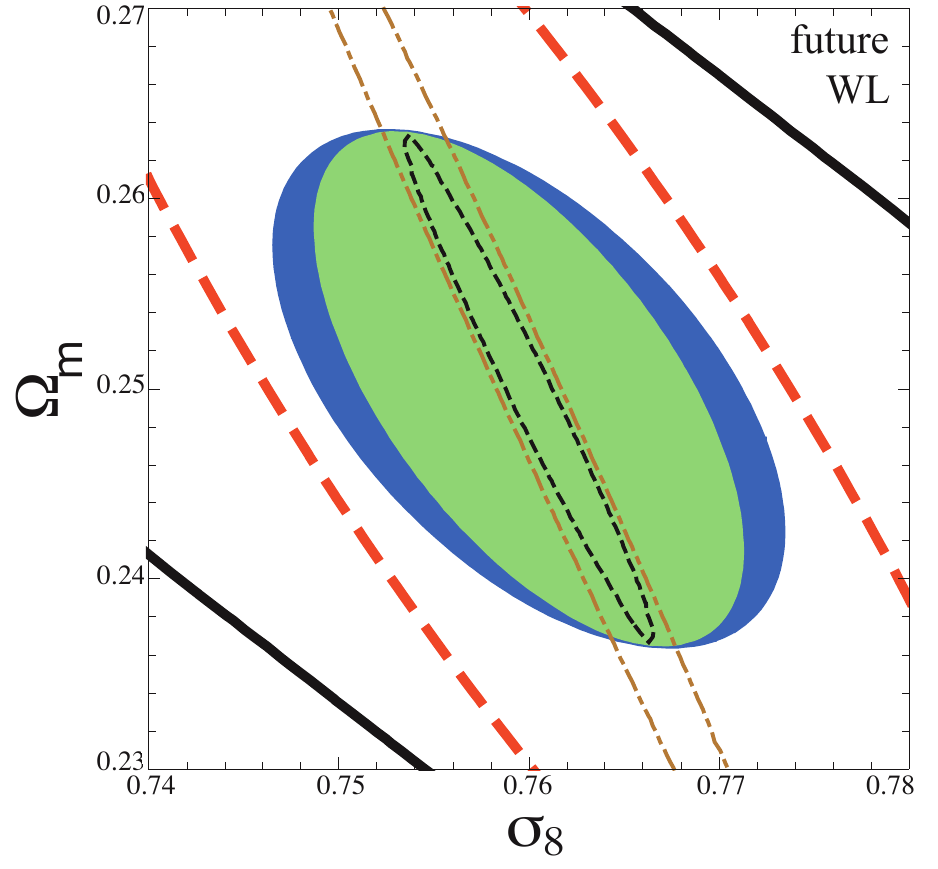}

\caption{Joint Constraints on $\Omega_{m}$ and $\sigma_8$ 
from SZ (top) and WL (bottom), near-future (left) and future (right) 
surveys. In all panels the fiducial model has $s_{eff} = 0$.
Legends are identical to Fig. 3
}
\label{DDE_SZ_Oms8_ce0$}
\end{figure}

Our main conclusions are:

\begin{quote}
1. Ongoing surveys such as the South Pole Telescope, 
which will detect a few thousands of clusters, are already
sensitive enough to detect dark energy perturbations (in the
absence of strong systematics) if $s_{eff} < 0$. 
Dark energy perturbations also weaken the detection of the other
dark energy parameters determined by the SPT, such as
$w_0$ and $w_a$: 
in particular, the marginalized figure-of-merit ($1/\sqrt{\Delta w_0 \Delta w_a}$)
is substantially smaller ($\sim$ 50\%)
compared to the case where dark energy
perturbations are neglected. This points to a
degeneracy between the effective sound 
speed and the equation of state which is not unexpected, since one determines 
the pressure on small scales, while the other sets the pressure on large scales.

2. Future surveys, such as the LSST, may or
may not find evidence
of dark energy perturbations. But even in the worst scenario
($s_{eff}>0$, weakest dependence on the sound speed) the constraints
on the main cluster parameters (such as $\Omega_m$ and $\sigma_8$)
are substantially weakened by this additional ``nuisance parameter". 
Furthermore, the marginalized limits on the equation of state parameters 
$w_0$ and $w_a$ are affected at the level of 10-30\%, depending on
the scenario -- see Figs. 3-4.

3. Surveys in the more distant future, such as ESA's Euclid or NASA's JDEM, 
will detect the imprints of the fluctuations of dark energy, but 
in the most conservative scenario the constraints would
only be strong enough to determine whether the dark energy pressure
in collapsed regions is positive or negative.

\end{quote}

Therefore, if future data
points to anything other than a Cosmological Constant, cluster surveys
will be a key observation to determine the nature of dark energy.
But even if cluster counts turn out to impose weak constraints on dark 
energy clustering, this would still be a minor ingredient affecting the halo 
mass function. Our results show that the addition variance (nuisance)
introduced by dark energy clustering
in the determination of the other cosmological parameters through
galaxy cluster counts cannot be dismissed.

\begin{acknowledgments}

R.A. would like to thank L. Liberato and M. Lima for useful discussions.
This work has been supported by FAPESP and CNPq of Brazil.

\end{acknowledgments}

\bibliographystyle{h-physrev3}
\bibliography{Letter_Fisher}

\end{document}